%% file: main.tex
\documentclass{article}

\usepackage[final,main]{neurips_2025}
\usepackage{hyperref}
\usepackage[utf8]{inputenc}
\usepackage[T1]{fontenc}
\usepackage{bm}
\usepackage{xspace}
\usepackage{amsmath}
\usepackage{amsfonts}
\usepackage[ruled,vlined]{algorithm2e}
\usepackage{physics}
\usepackage{graphicx}
\usepackage[caption=false]{subfig}
\usepackage{booktabs}
\usepackage{xcolor}
\usepackage[capitalise]{cleveref} 
\usepackage{verbatim}

\input{new_commands.tex}

\hypersetup{
    colorlinks=true,
    linkcolor=black,
    citecolor=purple,
    urlcolor=blue,
    }

\graphicspath{{./}{figures/}}

\title{DegenDetector: Symbolic Recovery of Parameter Degeneracies in Bayesian Posteriors}
\workshoptitle{2026 Conference on Physics and AI (PAI26)}
\begin{document}
\author{%
  Chaipat Tirapongprasert$^{1*}$ \quad
  Matthew Ho$^{1}$ \\
  $^{1}$\columbia
}

\maketitle

\begin{abstract}
We introduce \textsc{DegenDetector}, a framework for identifying and characterizing parameter degeneracies in posterior distributions as closed-form symbolic equations. By combining mutual information screening with alternating symbolic regression, we facilitate automated and interpretable identification of degenerate relationships without domain-specific input.
While standard tools such as corner plots can indicate that correlations exist, they do not reveal the underlying functional form.  \textsc{DegenDetector} fills this gap by expressing multi-parameter degeneracies as closed-form equations, providing interpretable structure that scales to high-order parameter spaces. 
\end{abstract}

\input{1_intro}
\input{2_methods}
\input{3_toy}

\input{4_science}

\section{Conclusion}
We developed \textsc{DegenDetector}, a novel degeneracy detection framework that integrates mutual information ranking with alternating symbolic regression to identify coupled parameters and render that degeneracy as a comprehensible equation. On synthetic benchmarks and Planck 2018 posteriors, we recover the true functional form of degeneracies with $R^2_\perp > 0.98$ across all cases.

The primary limitation in our method is the assumption of separability; future research could extend the framework to non-separable functional forms via multivariate symbolic regression or learned reparameterizations. All code is publicly available on Github.\footnote{\url{https://github.com/chaipattira/degen_detector}}
\newpage
\bibliography{references}
\end{document}

%% file: new_commands.tex
\newcommand{\columbia}{Department of Astronomy, Columbia University, New York, NY, 10027}

%% file: 1_intro.tex
\section{Introduction} \label{sec:intro}

A central goal in physics is to constrain the parameters of physical models using experimental data, which, within Bayesian statistics, amounts to computing a posterior
distribution over model parameters. Markov Chain Monte Carlo \citep[MCMC;][]{Goodman2010,Skilling2006} algorithms allow for obtaining such posteriors provided that we can write down the analytic likelihood.
Complex physical phenomena, however, often necessitate numerically intensive simulations to evaluate the likelihood. By training a neural surrogate to mimic simulation outputs, simulation-based inference \citep[SBI;][]{Papamakarios2019} has become the standard framework to address these intractable likelihoods.

Parameter degeneracy is a persistent challenge in both settings. This occurs when the observed data constrains a combination of parameters more tightly than the individual parameters themselves, causing posterior samples to concentrate along a lower-dimensional manifold. Some canonical cosmological examples include the CMB power spectrum \citep{Efstathiou1999} and weak-lensing degeneracies
\citep{Jain1997}. Left undetected, such degeneracies have serious downstream consequences, such as producing deceptively narrow error bars for coupled parameter combinations that explode for individual parameters \citep{Efstathiou1999}. MCMC marginalization over degenerate directions can also produce one-dimensional posteriors that misrepresent the best-fit region entirely, yielding confidence intervals that are artifacts of the projection rather than reflections of the data themselves \citep{colgain2025}.

Detecting degeneracies has proved to be a difficult task \citep{Jasche2013,Fluri2021}, especially for a joint distribution of more than two parameters, as these high-order degeneracies do not manifest in the pairwise projections given by corner plots and other standard diagnostic tools.  
To address these challenges, we present \textsc{DegenDetector}, an automated pipeline that (a) identifies which subsets of parameters contribute to the degeneracy and (b) expresses such degeneracy as a closed-form symbolic equation that can be used to inform reparameterization \citep{villa2025addressingpriordependencehierarchical} and decorrelate the degeneracy.
We demonstrate our algorithm on synthetic posteriors with known analytical degeneracies and apply it
to posterior chains from Planck CMB measurements \citep{Planck2020}.

%% file: 2_methods.tex
\section{Methods} \label{sec:methods}

Consider a Bayesian inference problem where we have used MCMC or neural SBI methods to generate $N$ samples $\{\boldsymbol{\theta}^{(i)}\}_{i=1}^N$ from a posterior distribution $p(\boldsymbol{\theta}\mid\mathbf{d})$ over an $M$-dimensional parameter space $\boldsymbol{\theta} = (\theta_1, \theta_2, \ldots, \theta_M)$.

\textbf{Degeneracies}\,\,
The samples are said to exhibit $k$-parameter degeneracy if there exists a smooth function
$F(\theta_{j_1}, \ldots, \theta_{j_k})$ such that  most posterior mass lies within an $\varepsilon$-neighborhood of
\begin{equation}
    \mathcal{M}_c = \bigl\{\,\boldsymbol{\theta} \in \mathbb{R}^M : F(\theta_{j_1}, \ldots, \theta_{j_k}) = c\,\bigr\},
\end{equation}
under some metric on parameter space, where $(j_1,\dots, j_k) \subseteq [1,M]$ is a tuple of the indices of the $k$ parameters involved in the degeneracy.
Generally, the level set $\mathcal{M}_c$ is a $(k-1)$-dimensional submanifold of $\mathbb{R}^M$, rendering parameter inference challenging due to the inability to resolve individual parameter values along degenerate directions.
Without knowledge of the degeneracy, we would not be able to predict how a change in one parameter would result in the same observations.

\subsection{Pinpointing Candidate Parameters}
\textbf{Mutual Information}\,\,
To find degeneracies, we need to narrow down the parameter combinations that could cause combinatorial explosions as $\binom{M}{k}$. \textsc{DegenDetector} uses mutual information (MI) to rank tuples based on their statistical dependence. Recall that MI of a parameter pair $I(\theta_a;\,\theta_b) = \int p(\theta_a,\theta_b)\log\frac{p(\theta_a,\theta_b)}{p(\theta_a)\,p(\theta_b)}\,\dd\theta_a\,\dd\theta_b$ measures the reduction in uncertainty about $\theta_a$ given knowledge of $\theta_b$.
 If $\theta_a$ and $\theta_b$ are statistically independent, MI is zero; otherwise, it is strictly positive.
 
To estimate the pairwise MI scores, we use the $k$-nearest-neighbor estimator
\citep{Kraskov_2004} and construct an $M\times M$ matrix that is subsequently symmetrized and floored at zero. MI is sensitive to arbitrary nonlinear dependence structures between parameters, making it well suited to capture the nonlinear, multi-valued relationship responsible for physically motivated degeneracies.

For a target coupling depth of $k$, we enumerate all parameter tuples $(\theta_{j_1},\ldots,\theta_{j_k})$, assign them a score equal to the sum of all pairwise MI values among their elements, and arrange them in descending order. Higher aggregated MI implies a stronger, more significant degeneracy.

\subsection{Symbolic Surface Fitting}

We model the degeneracy as the level set of a separable function,
\begin{equation}
    g_1(\theta_{j_1}) + g_2(\theta_{j_2}) + \cdots + g_k(\theta_{j_k}) = c\, ,
\end{equation}
where $g_l$'s are univariate functions to be discovered and $c$ is a
real constant.

\textbf{Alternating Optimization}\,\, \textsc{DegenDetector} fits the component functions $\{g_\ell\}_{\ell=1}^{k}$ and the constant $c$ using alternating optimization, cycling through one component at a time while holding the others constant.
This allows us to separate the $k$-dimensional fitting problem into $k$ independent one-dimensional symbolic regression problems, resulting in a faster, more effective search. In what follows, we will use $\theta_\ell$ as a shorthand for $\theta_{j_\ell}$ for a fixed candidate tuple $(j_1,\ldots,j_k)$.

To ensure that parameters with different units or physical scales contribute equally, we first normalize each parameter using the z-score $\tilde{\theta}_\ell = \frac{\theta_\ell - \mu_\ell}{\sigma_\ell}$,
where $\mu_\ell$ and $\sigma_\ell$ are the sample mean and standard deviation.
Each component function is then initialized to the identity, $g_\ell(\tilde\theta_\ell) = \tilde\theta_\ell$,
and the constant to $c = \mathrm{mean}\bigl(\sum_\ell \tilde\theta_\ell\bigr)$.
At each alternating step, we cycle through all $k$ components and establish the regression target: 
    $y_\ell = c - \sum_{m\neq \ell} g_m(\tilde\theta_m)$,
for component $\ell$, which is what
$g_\ell(\tilde\theta_\ell)$ ought to produce for every sample if the constraint is
satisfied.  We then perform symbolic regression on the pairs
$\{(g_\ell(\tilde\theta_\ell)^{(i)},\, y_\ell^{(i)})\}_{i=1}^{N}$ using \textsc{PySR}
\citep{Cranmer2023}.
Once convergence is achieved, we use \textsc{SymPy} \citep{meurer2017sympy} to simplify and restore each component to the original scale, yielding a human-readable equation for the degeneracy.

\textbf{Orthogonal Fit Quality}\,\,
For an implicit surface $F(\boldsymbol{\theta}) = \sum_\ell g_\ell(\theta_\ell) - c = 0$, the ordinary coefficient of determination $R^2$ is ill-defined due to its asymmetry with respect to component function permutations (i.e., one parameter is specifically chosen to serve as the dependent variable).
We define the orthogonal $R^2$ as a more appropriate goodness of fit metric $R^2_\perp = 1 - \mathcal{L}_\perp$
with the orthogonal loss being the mean squared perpendicular distance from the posterior samples to the fitted surface,
\begin{equation}
    \mathcal{L}_\perp = \frac{1}{N}\sum_{i=1}^{N}\frac{\bigl[F(\boldsymbol{\theta}^{(i)})\bigr]^2}{\|\nabla F(\boldsymbol{\theta}^{(i)})\|^2} = \frac{1}{N}\sum_{i=1}^{N} \frac{\bigl[\sum_\ell g_\ell(\theta_\ell^{(i)}) - c\bigr]^2}{\sum_\ell \bigl[g_\ell'(\theta_\ell^{(i)})\bigr]^2 + \varepsilon}\,.
\end{equation}

When true degeneracy is not separable, the orthogonal $R^2$ will return a low score, indicating that the functional form has not been captured.
The scientist can discover multiplicative degeneracies in log-space by activating \textsc{LogDegen}, which applies a coordinate transformation $\tilde{\theta}_i = \log\theta_i$, reruns the pipeline on the transformed coordinates, and reports back in the original parameterization.

\textbf{Diagnostics} \,\, For each fitted tuple, \textsc{DegenDetector} provides a ranking of the top 5 candidate symbolic
equations together with their $R^2_\perp$ scores and expression complexities.
A suite of diagnostic visualizations is also generated: a heatmap of
the pairwise MI matrix identifying the most coupled parameter pairs, a corner
plot of the degenerate parameters with the fitted surface overlaid as a contour,
true-versus-predicted plots for each component function, and two- or three-dimensional renderings of the fitted manifold.  These
diagnostics allow the scientist to assess whether the separable model
captures the true degeneracy structure before accepting the symbolic equation.

%% file: 3_toy.tex
\section{Benchmark Examples} \label{sec:toy}

\begin{figure}[t]
    \centering
    \includegraphics[width=\linewidth]{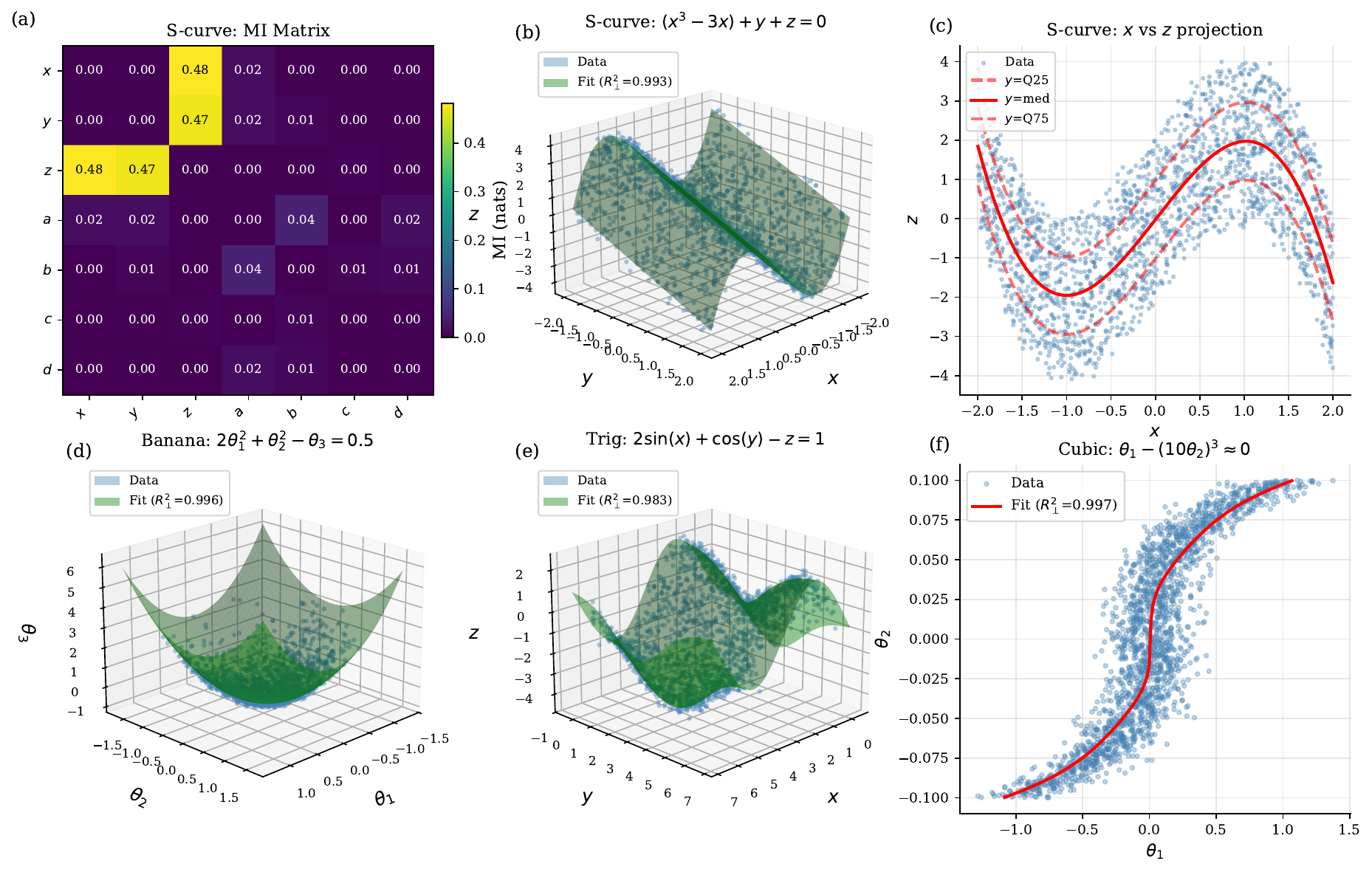}
    \caption{\textbf{Validation of \textsc{DegenDetector} on four benchmarks.}
    \textit{Top row} shows the full pipeline for the S-curve degeneracy.
    \textit{(a)}~Pairwise MI matrix: the degenerate parameter tuple $(x,y,z)$ forms a high MI cluster with pairwise MI $\approx 0.47$--$0.48$, while nuisance parameters are cleanly separated.
    \textit{(b)}~Fitted surface (green) recovers the posterior samples with $R^2_\perp = 0.9931$.
    \textit{(c)}~Two-dimensional $x$--$z$ projection of the same fit.
    \textit{(d-f)} shows fitted manifolds for the remaining three experiments.}
    \label{fig:toy_combined}
\end{figure}

We validate \textsc{DegenDetector} on synthetic posteriors that contain a priori defined, strong,
non-Gaussian degeneracies along with some independent nuisance parameters (drawn from an isotropic Gaussian distribution).
For the polynomial regime, we begin with the Rosenbrock function, a standard benchmark problem in numerical optimization ~\citep{Goodman2010} and extend the characteristic \textit{banana}-shape constraints onto three dimensions. We then proceed to a more complex, non-monotone degeneracy in \textit{S-curve}. In \textit{Trig}, we assess if our method can approximate periodic functions with a limited set of polynomial-style operators, possibly through Taylor expansions or Gaussian-envelope estimations, and still achieve sufficient accuracy for the diagnostic to be useful. Finally, in \textit{Cubic}, we test on posteriors sampled from an anisotropic prior with mismatched parameter scales, which ubiquitously occurs when physical parameters have different units. 

Figure \ref{fig:toy_combined} illustrates how \textsc{DegenDetector} successfully recovers these non-trivial degeneracy across the four experiments.
We find that \textsc{DegenDetector} can distinguish the underlying degenerate parameters from the other $\binom{7}{3} = 35$ candidate tuples and recover an acceptable functional form (with $R^2_\perp > 0.98$).

%% file: 4_science.tex
\section{Science Experiment}

We apply \textsc{DegenDetector} to posterior chains from the
Planck 2018 baseline analysis \citep{Planck2020} with seven cosmological parameters $\Omega_b h^2$,
$\Omega_c h^2$, $H_0$, $\Omega_m$, $\sigma_8$, $n_s$, and $\tau$. The CMB horizon-angle degeneracy $\Omega_m h^3 \approx \mathrm{const}$ is well established; the exponent of 3 \citep[][empirically find $\sim 3.4$]{Percival2002} reflects the combined
dependence of the angular scale of the
sound horizon at last scattering on the matter density and expansion rate \citep{Efstathiou1999}. Although the power law degeneracy is non-separable in the $(\Omega_m, h)$ space, it can be recovered as the level set $\log\Omega_m + 3\log h = \mathrm{const}$ in $(\log\Omega_m,\,\log H_0)$ space.

\begin{figure}[t]
    \centering
    \includegraphics[width=\linewidth]{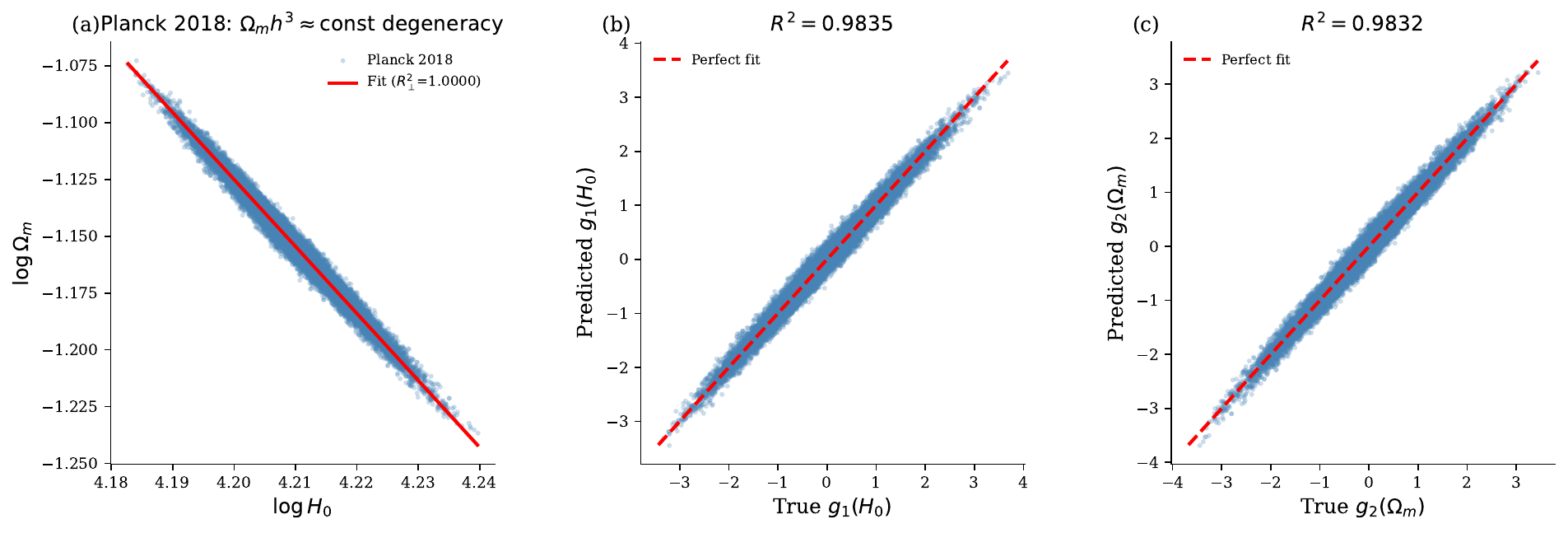}
    \caption{\textbf{Automated recovery of the CMB horizon-angle degeneracy from Planck 2018 posterior samples.}
    \textit{(a)}~Scatter of 25{,}225 Planck TTTEEE+lowl+lowE+lensing samples in log-parameter space,
    overlaid with the fitted degeneracy line (red).
    The tight linear locus confirms that $\Omega_m h^3 \approx \mathrm{const}$ is the dominant
    constraint.
    \textit{(b--c)}~True vs.\ predicted values for each component function $g_1(H_0)$ and
    $g_2(\Omega_m)$, demonstrating that both are well described by simple log-linear forms.}
    \label{fig:planck_h0_omegam}
\end{figure}

Without any prior knowledge of the physics, \textsc{LogDegen} recovers $(H_0,\,\Omega_m)$,
with a mutual-information score of $I = 2.07$ and fits the
Planck posterior in log-space, returning the expression:
\begin{equation}
  123.97\,\log H_0 + 42.07\,\log\Omega_m = \mathrm{const}\, ,
  \label{eq:detected}
\end{equation}
with $R^2_{\mathrm{\perp}} \approx 0.98$ and a residual standard deviation of
$\sigma_\perp = 0.128$ in log-space.  Dividing by the $\Omega_m$ coefficient yields a ratio of $123.97 / 42.07 \approx 2.947$, which is reasonably consistent with the anticipated functional form up to a constant.
The manifold fit and its agreement with the posterior
samples are illustrated in Figure~\ref{fig:planck_h0_omegam}.